\documentclass[mypaper,7pt,twoside]{CoAst}
\usepackage{epsf,graphicx,fancyhdr}
\renewcommand{\chaptermark}[1]{\markboth{#1}{}}

\renewcommand{\headrulewidth}{0pt}
\newcommand{\apj}{ApJ}
\newcommand{\aap}{A\&A}  
\newcommand{\mnras}{MNRAS}

\newcommand{\kopf}{\small\itshape Comm. in Asteroseismology\\ Vol. 143, 2003}
\newcommand{\Authors}[1]{\begin{center}\normalsize\bf\sf #1 \end{center}}

\renewcommand{\author}[1]{\begin{center}\normalsize\bf\sf #1 \end{center}}
\newcommand{\Address}[1]{\begin{center}\small\sf #1 \end{center}}

\renewenvironment{abstract}{\section*{Abstract}\normalsize\sf}{}
\newcommand{\References}[1]{\begin{flushleft}{\large References\\}\vspace*{2mm}\small #1 \end{flushleft}}

\newcommand{\chapterDSSN}[2]{\chapter[\sf\normalsize #1\\ \footnotesize \hspace*{5mm}by #2 \sf\normalsize][]{#1\\}\rhead[\fancyplain{}{\sf\footnotesize \center{#1}}]{\fancyplain{}{\sffamily\thepage}}\lhead[\fancyplain{\kopf}{\sffamily\thepage}]{\fancyplain{\kopf}{\sf\footnotesize \center{#2}}}}

\renewcommand{\chaptername}{}

\newcommand{\acknowledgements}[1]{\vspace*{5mm}\noindent\begin{bf}Acknowledgments. \end{bf} #1}

\def\lesssim{\mathrel{\hbox{\rlap{\hbox{\lower4pt\hbox{$\sim$}}}\hbox{$<$}}}}

\begin{document}
\sf

\chapterDSSN{$\it{MOST}$ Photometry of the roAp star\\ HD 134214 }{C.
Cameron et al.}

\Authors{C. Cameron$^1$, J.M. Matthews$^1$, J.F. Rowe$^1$, R.
Kuschnig$^1$, D.B.
Guenther$^2$,
  A.F.J. Moffat$^3$, S.M. Rucinski$^4$, D. Sasselov$^5$, G.A.H.
Walker$^6$, W.W.
Weiss$^7$}
\Address{
$^1$Dept. of Physics and Astronomy, UBC, 6224 Agricultural Road,\\ Vancouver,
BC, V6T 1Z1, Canada\\
$^2$Dept. of Astronomy and Physics, Saint Mary's University,\\
Halifax, NS, B3H 3C3, Canada\\
$^3$D\'ept. de physique, Univ. de Montr\'eal C.P.\ 6128, Succ. Centre-Ville,\\
Montr\'eal, QC H3C 3J7, Canada; and Obs. du Mont M\'egantic\\
$^4$Dept. of Astronomy \& Astrophysics, David Dunlap Obs., Univ. Toronto
P.O.~Box 360, Richmond Hill, ON L4C 4Y6, Canada\\
$^5$Harvard-Smithsonian Center for Astrophysics, 60 Garden Street,\\
Cambridge, MA 02138, USA\\
$^6$1234 Hewlett Place, Victoria, BC V8S 4P7, Canada\\
$^7$Institut f\"ur Astronomie, Universit\"at Wien T\"urkenschanzstrasse 17,\\ A--1180 Wien, Austria}

\noindent
\begin{abstract}
We present 10.27$~\rm{hrs}$ of photometry of the roAp star HD 134214
obtained by the $\it{MOST}$\footnote{$\it{MOST}$ (Microvariability
and Oscillations of STars) is a Canadian Space Agency mission, operated
jointly by Dynacon, Inc., and the Universities of Toronto and British
Columbia, with assistance from the University of Vienna.} satellite.
The star is shown to be monoperiodic and oscillating at a frequency of
2948.97$~\pm~$0.55$~\mu\rm{Hz}$. This is consistent with earlier
ground based photometric campaigns (e.g. Kreidl et al. 1994).  We do
not detect any of the additional frequencies identified in the recent
spectroscopic study by Kurtz et al. (2006) down to an amplitude limit
of 0.36$~\rm{mmag}$ ($2\sigma$ significance limit).
\end{abstract}

\section{1. Introduction}

The rapidly oscillating Ap stars (roAp) represent a unique subset of
the chemically peculiar stars of the upper main sequence. In general,
the Ap stars have globally organised magnetic fields of strengths of
order kiloGauss, and spectral anomalies that are interpreted as vertical
and horizontal chemical inhomogeneities in the stellar atmosphere.
The roAp stars (among the coolest members of the Ap class) exhibit
rapid oscillations in photometry and spectroscopy. These variations
(first observed by Kurtz (1982))  have periods from about 5 to 20
minutes and low amplitudes (B~$\lesssim$~10 mmag).  They are consistent
with acoustic (p-mode) pulsations of low degree and high radial overtone.
A thorough review of the roAp stars is provided by Kurtz \& Martinez
(2000).

HD 134214 is in many ways a typical Ap star. It has an effective
temperature of $\sim$ 7500 K, is a moderately slow rotator with a
$v \sin i \approx 2.0$ km/s, and has a strong magnetic field with $B_{r}
\approx -2.9$ kG (parameters are derived in Shavrina et
al. 2004). However, among the roAp stars, HD 134214 stands alone as
the star with the shortest known period ($\sim 5.6$ minutes). This
high frequency oscillation is well above the estimated isothermal
acoustic cutoff for a typical A-star model (see, e.g., Audard et
al. 1998). There is also evidence that this oscillation frequency is
variable at the level of a few tenths of$~\mu\rm{Hz}$ over a time
scale of approximately one year, even though the amplitude remained
stable over many years (see Kreidl et al. 1994). Kurtz (1995) discuss
the frequency variability observed in this and other roAp stars.

Recently, Kurtz et al. (2006) have shown that in some roAp stars the
spectroscopic variability is dramatically different from that observed
photometrically. Although HD 134214 is observed to be monoperiodic in
photometry over many years (Kreidl et al. 1994), Kurtz et al. present
evidence for up to 6 periodicities in their (admittedly short)
spectroscopic time-series of this star.

In this paper we present $\it{MOST}$ (Microvariability \& Oscillations
of STars) spacebased photometry of HD 134214 and compare the observed
oscillations to the results of Kurtz et al. (2006).

\section{2. Photometry and  Frequency Analysis}\label{sec2}

$\it{MOST}$ is a Canadian microsatellite that was launched into a 820-km
Sun-synchronous polar orbit in June 2003. Its primary science objective
is to obtain nearly continuous, ultra-precise photometric measurements
of stars for the purposes of asteroseismology. The instrument is a 15-cm
Rumak-Maksutov telescope that illuminates a CCD photometer through a
custom broadband filter (350-700 nm). Detailed technical information on
the instrument is contained in Walker, Matthews et al. (2003). The first
science results for the mission were published by Matthews et al. (2004).

The roAp star HD 134214 was observed by $\it{MOST}$ as a trial Direct
Imaging target on 1 May 2006. Direct Imaging is the observing mode where
a defocused star image is projected onto an open area of the Science CCD
(see Rowe et al. 2006; and references therein, in this issue for
more details on different observing modes of $\it{MOST}$). The star
was in one of two target fields observed during each $\it{MOST}$ satellite
orbit (period = 101.413 min), so data are collected with a duty cycle
of only $\sim$ 47\%.  Exposures are 1.5 sec long, obtained every 10
seconds. There were a total of 505 measurements collected during 10.27 hrs.

Photometric reductions for Direct Imaging targets are described by Rowe et
al. (2006b). In addition to these procedures, the data are detrended using
a running mean of approximately 15 mins to reduce any remaining
low-frequency contributions to stray light from scattered Earthshine which
is modulated with the $\it{MOST}$ satellite orbital period. This detrending does
not influence the high-frequency domain of interest for roAp
oscillations. The resulting changes in the mean level of the light
curve are not large enough to affect the measured amplitude of the
oscillation signal, within the uncertainty from the fit (see below).

\begin{figure}[!h]
\centering
\includegraphics[scale=.65,clip]{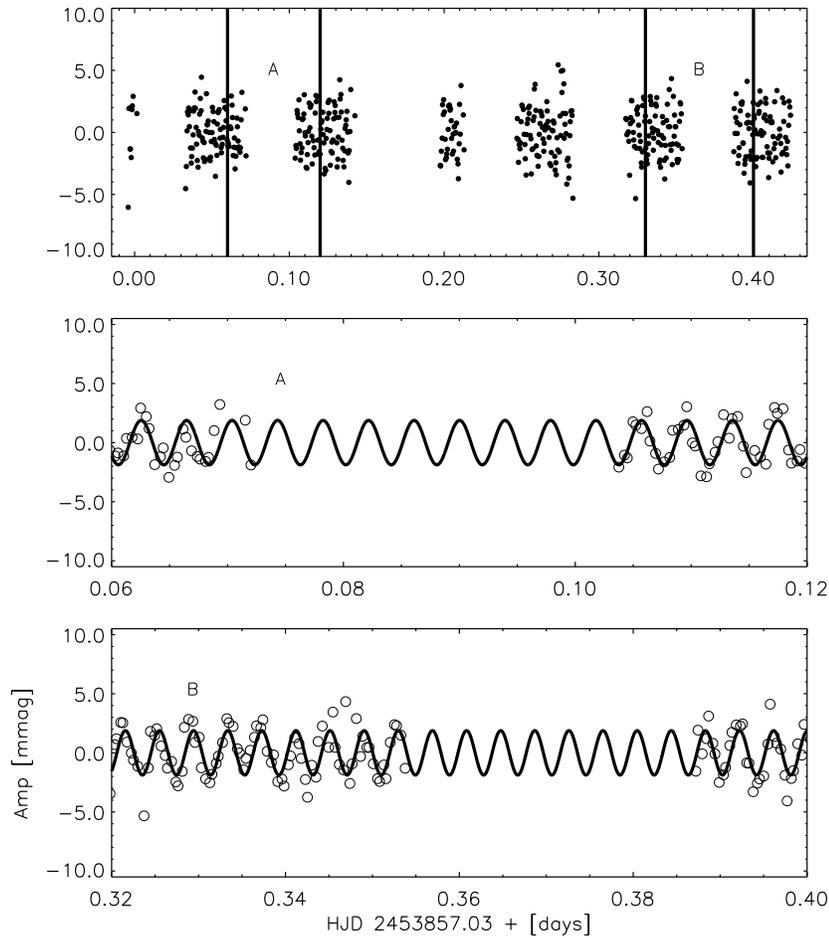}
\caption[Figure 1:]{The $\it{MOST}$ photometry of HD 134214. The top
  panel shows the 10.27 $\rm{hrs}$ of data. Gaps in the light
  curve are the result of observing this target every second satellite
  orbit. The zoomed regions; labelled A and B, are shown
  respectively in the lower panels. The open circles are approximately
  twice the standard error in size ($\sim 0.20$ mmag). The solid line
  represents the fit of a single sinusoid with a frequency of
  2948.97$~\mu\rm{Hz}$}
\label{Fig1}
\end{figure}

The reduced photometry is presented in Figure 1. The light curve of
HD 134214, shown in the top panel, has a standard deviation of $\sim
1.8$ mmag and a $2\sigma$ standard error of 0.2 mmag. Data subsets
labelled A and B are shown in the lower two panels.

Frequency determinations are made using CAPER; see Walker et al. (2005)
and Saio et al. (2006). CAPER is an iterative procedure that identifies
periodicities in Fourier space and simultaneously fits a set of
sinusoids in the time domain. This method of determining oscillation
parameters from time-series data follows the philosophy of the popular
software packages Period98 (Sperl 1998) and Period04 (Lenz \& Breger
2005). Discrete Fourier Transforms (DFTs) are calculated and
the fit is subtracted from the data successively, until there is no
meaningful change in the fit residuals.

\begin{figure}[!ht]
\centering
\includegraphics[scale=0.428,clip]{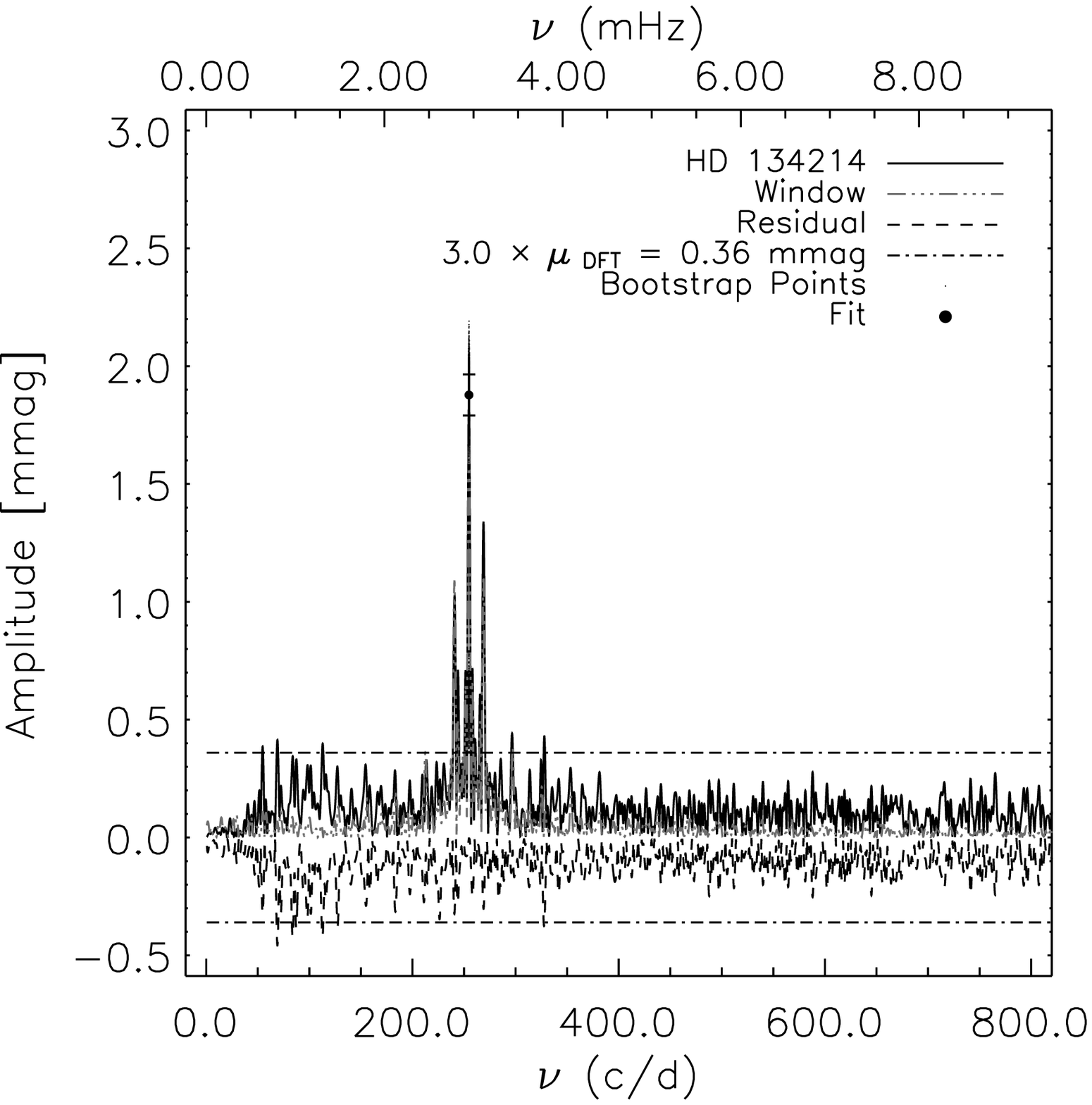}\\
\includegraphics[scale=0.428,clip]{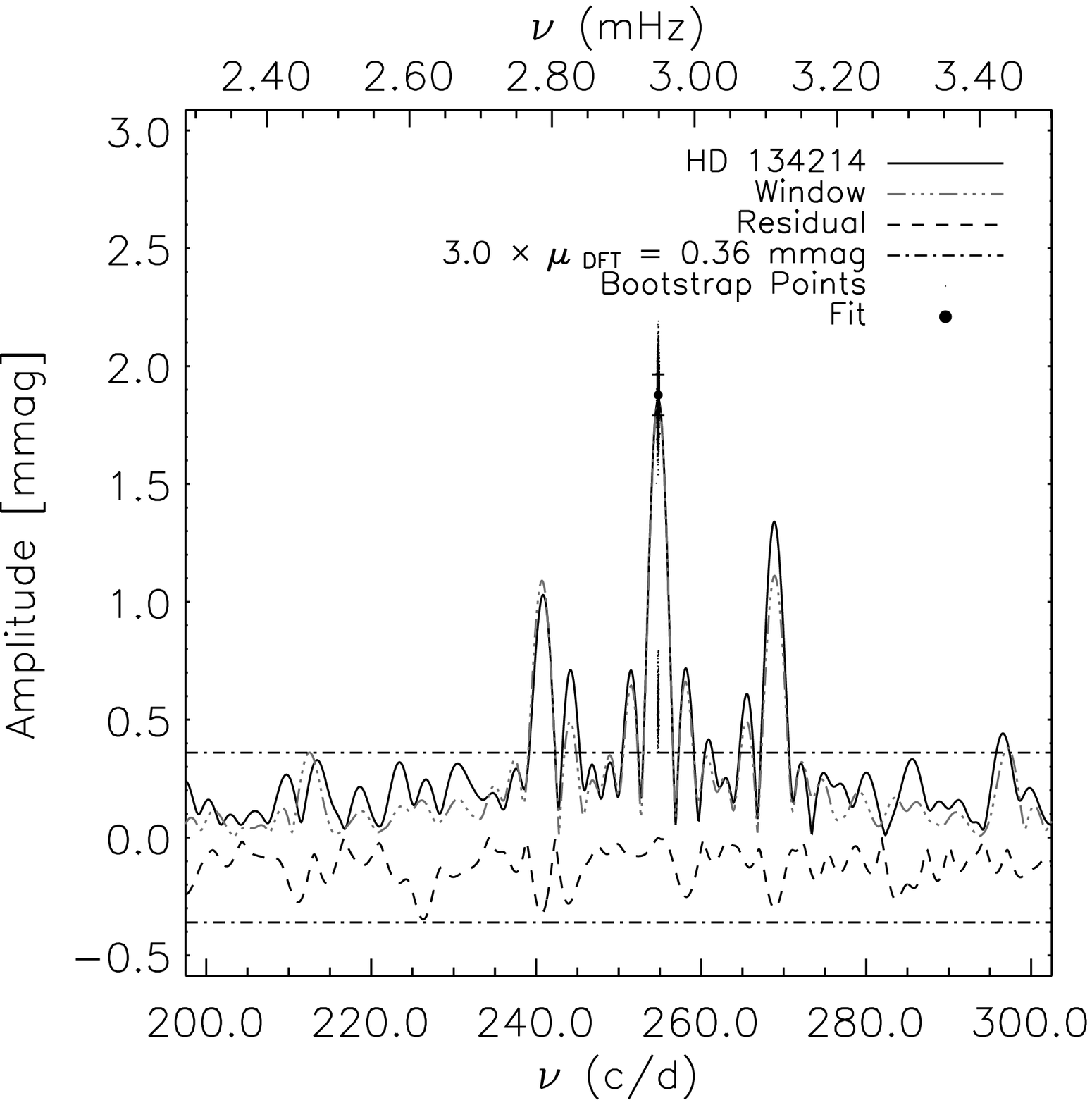}
\caption[Figure 2:]{a) (top) The DFT of the HD 134214 data. Shown
are the window function, the fit to the data (filled circle) and the
bootstrap points (small dots). Also shown is the residual DFT after
the fit; inverted for clarity. b) (bottom) The DFT of the data zoomed
in on the region of the largest peak. Line styles are the same as in
the top plot.}\label{Fig2}
\end{figure}

The DFT and spectral window function for the HD 134214 data are shown in
Figure 2. There is one periodicity identified at
2948.97$~\pm~$0.55$~\mu\rm{Hz}$
with a signal-to-noise ratio of 15.4. The noise in the Fourier domain was
estimated as the mean of the entire spectrum (0.12 mmag). We see no evidence
for other periodicities at a detection threshold of about $2\sigma$ (or
$3\times$ the noise; see Kuschnig et al. 1997). The fitted values
of amplitude and phase are 1.88$~\pm~$0.09$~$mmag and 1.97$~\pm~$0.39$~$
radians respectively. The uncertainties in the fit parameters are determined
using a bootstrap technique (see below).  The dark line in Figure 1 shows
the fit of this single sinusoid plotted over the observations.

\begin{figure}[!h]
\centering
\includegraphics[scale=.5,clip]{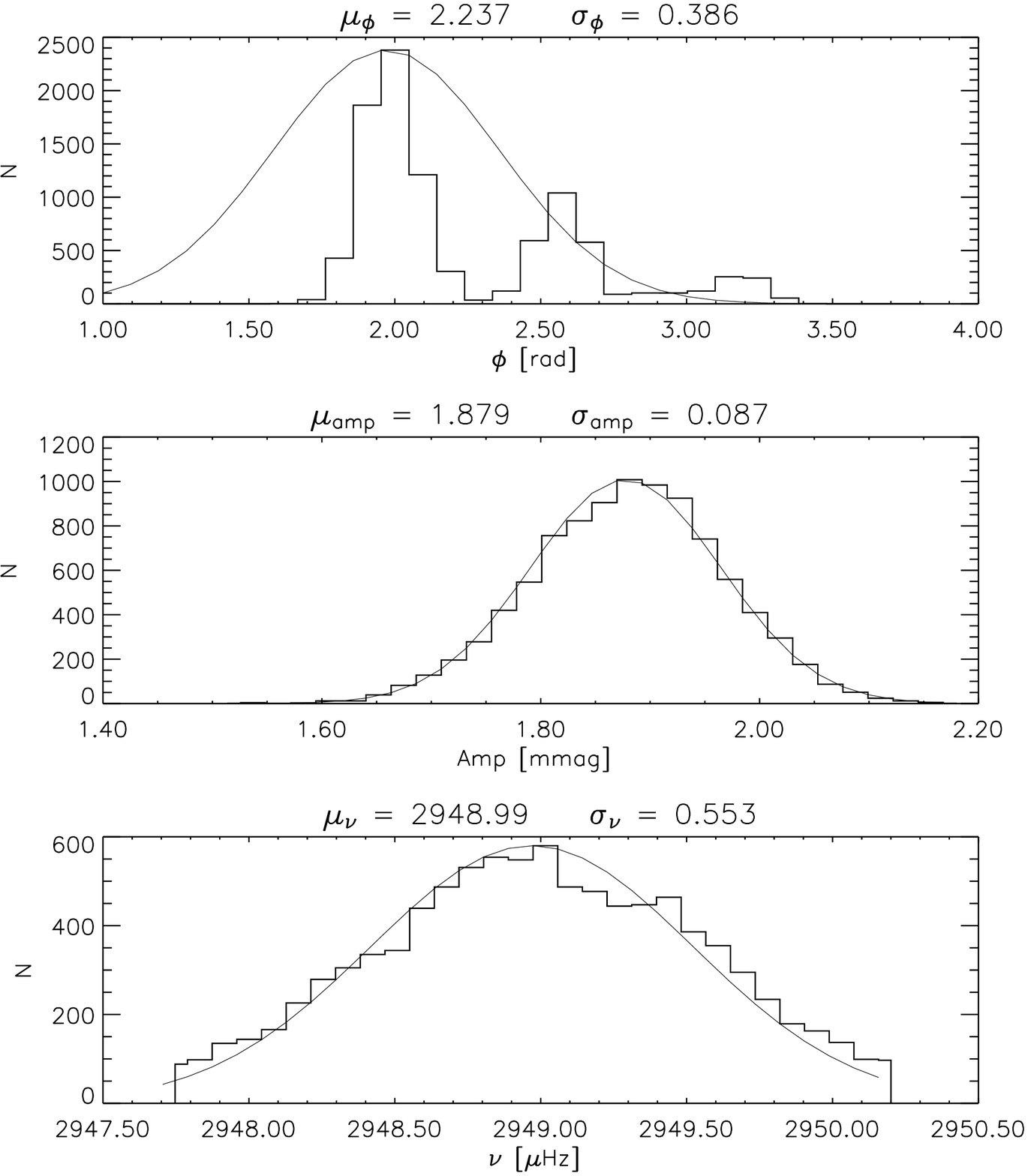}
\caption[Figure 3:]{The bootstrap distributions for the fit
parameters. Shown from top to bottom are the distributions
for the phase $\phi$ in radians, the amplitude in mmag, and the
frequency $\nu$ in $\mu\rm{Hz}$. The labels above each panel give the
mean $\mu$ and the standard deviation $\sigma$ calculated for each
distribution. Gaussians with width $\sigma$ and centered on the
original fit parameters are plotted over each distribution.}\label{Fig3}
\end{figure}

The uncertainties we calculate for time-series parameters derived from
nonlinear least-squares methods depend on the noise of the data (which may
be a combination of instrumental and random processes) and on the time
sampling. It is also known that fitted phase and frequency parameters are
correlated, leading to underestimated uncertainties in these parameters \
when calculated from a covariance matrix (see, e.g., Montgomery \&
O'Donoghue 1999).

The "bootstrap" is an ideal way to assess the uncertainties in fitted
parameters for a time-series analysis. The procedure is outlined in
Wall \& Jenkins (2003) and has recently been used in a number of
$\it{MOST}$ applications (see, e.g., Rowe et al. 2006b and Saio
et al. 2006). Clement et al. (1992) also used a bootstrap to estimate
the uncertainties in Fourier parameters they derived for RR Lyrae stars.

In short, the bootstrap is a technique that allows a user to produce a
distribution for each calculated parameter by constructing a large number
of light curves from the original data. No assumptions need to be made
about noise properties of the data and individual photometric errors are not
needed. Each new light curve is assembled by randomly selecting N points
from the original light curve (also containing N points) with the
possibility of replacement. In this way, the new synthetic light curves
preserve the noise properties of the original data. The fit procedure is
repeated for each new light curve, eventually building distributions in
each of the fit parameters. We then estimate the 1$\sigma$ error bars
from the analytic expression for the standard deviation of each
distribution under the assumption that they are normally distributed.
Each distribution is checked to ensure this assumption is valid.

Figure 3 shows the bootstrap distributions for 10,000 realisations of the
fit to our data. The phase is the most uncertain parameter. Amplitude
and frequency distributions are normal in shape. Overplotted are
Gaussians with the standard deviations derived from the distributions.
The Gaussians are centred on the original fit parameters. It should be
stressed that bootstrapping only estimates the uncertainties in
parameters and does not refine them or assign significances to them.

\section{3. Discussion}

We determine that HD 134214 is monoperiodic and pulsating with a
frequency of 2948.97$~\pm~$0.55$~\mu\rm{Hz}$. This is consistent
with the earlier photometric observations of this star by Kreidl et
al (1994). Those authors detected frequency changes which they
interpreted as cyclic with a timescale of about 1 year. The frequency
resolution of our short $\it{MOST}$ run is insufficient to rigorously
test this assertion. However, we have considered the implications of 
pulsational frequency variability in HD 134214.

Heller \& Kawaler (1988) have suggested that it may be possible
to exploit the frequency variability of roAp stars to determine
their evolutionary status.  However, the one-year timescale of
variability in the case of HD 134214 is too short to be associated
with evolution.  We have also modelled the amplitude of the frequency
change over the main sequence life time of an A-type star. We estimate
that a 20 year baseline of observations would show a frequency
change on the order of $10^{-4} - 10^{-5}$$~\mu\rm{Hz}$. 
This is much smaller than the change of $\sim0.2~\mu\rm{Hz}$ reported
by Kreidl et al (1994). Cameron et al. (in preparation) have shown
that taking magnetic perturbations to the oscillation frequencies into
account does not improve the discrepancy between the observed and
calculated frequency changes.

Recent results by Kurtz et al. (2006) suggest that HD 134214 is
oscillating spectroscopically in up to 6 modes. Our observations
can rule out additional photometric oscillations at a $2\sigma$
detection level of 0.36 mmag. Our estimated DFT noise level of
$\sim$ 0.12 mmag for 10 hours of observations with $\it{MOST}$
is comparable to that obtained by Kreidl et al. (1994) based
on about 56 hours of photometry from four observatories during
about 4 months. The excitation and selection of pulsation modes
in roAp stars is an open question. In the case of HD 134214,
there are several avenues to be explored with respect to the
additional modes seen spectroscopically: (1) Is it a case of
radial velocities of certain elements and ionisation stages
being more sensitive to degrees of higher $\ell$? (2) If the
sensitivity of the photometry can be improved sufficiently,
will these modes also appear? (3) Is the broadband photometry
just averaging over too wide an extent of the pulsating
atmosphere? or (4) Might there be new physics in the upper
atmospheres of (some) roAp stars to account for the differences
observed?

We estimate that even observations over only a few days with
$\it{MOST}$ would (in addition to improving our frequency
resolution enough to investigate cyclic variability) reduce
our noise levels by approximately half. Even this modest
improvement in noise would be valuable as a test of the Kurtz
et al. (2006) theory that photometric observations are not
sensitive to the new type of upper atmosphere pulsational
variability they report.

\acknowledgements{JMM, DBG, AFJM, SR, and GAHW are supported by
funding from the Natural Sciences and Engineering Research Council
(NSERC) Canada.  RK is funded by the Canadian Space Agency. WWW
received financial support from the Austrian Science Promotion
Agency (FFG - MOST) and the Austrian Science Funds (FWF - P17580).}

\References{
Audard, N., et al.\ 1998, \aap, 335, 954\\
Cameron, C. et al. (in preparation)\\  
Clement, C.~M., Jankulak, M., \& Simon, N.~R.\ 1992, \apj, 395, 192\\
Heller, C.~H., \& Kawaler, S.~D.\ 1988, \apj, 329, L43\\ 
Kreidl, T.~J., et al.\ 1994, \mnras, 270, 115\\
Kurtz, D.~W.\ 1982, \mnras, 200, 807\\
Kurtz, D.~W.\ 1995, ASP Conf.~Ser.~ 76: GONG 1994.~Helio- and Astro-Seismology from the Earth and Space, 76, 60\\
Kurtz, D.~W., \& Martinez, P.\ 2000, Baltic Astronomy, 9, 253\\
Kurtz, D.~W., Elkin, V.~G., \& Mathys, G.\ 2006, \mnras, 370, 1274\\
Kuschnig, R., et al.\ 1997,\aap,328, 544\\
Lenz, P., \& Breger, M.\ 2005, Communications in Asteroseismology, 146, 53\\
Matthews, J.~M., et al.\ 2004, Nature, 430, 51\\
Montgomery, M.~H., \& O'Donoghue, D.\ 1999, DSSN (Vienna), 13\\
Rowe, J.~F., et al.\ 2006, CoAST (this issue)\\
Rowe, J.~F., et al.\ 2006b, \apj, 646, 1241\\
Saio, H., et al.\ 2006, \apj,  in press\\
Shavrina, A., et al.\ 2004, IAU Symposium, 224, 711\\
Sperl, M.\ 1998, Communications in Asteroseismology, 111, 1\\
Walker, G., et al.\ 2003, PASP, 115, 1023\\
Walker, G., et al.\ 2005, \apj, 635, L77\\
Wall, J.~V., \& Jenkins, C.~R. 2003, Practical Statistics for Astronomers,Cambridge,Cambridge University Press\\

}

\end{document}